\documentstyle[prl,aps,floats]{revtex}
\begin{document}
\renewcommand{\topfraction}{0.8}
\twocolumn[\hsize\textwidth\columnwidth\hsize\csname
@twocolumnfalse\endcsname

\title{A time-space varying speed of light
       and the Hubble Law in static Universe}
\author{Sergey S. Stepanov \\
Dnepropetrovsk State University\\
E-MAIL: steps@tiv.dp.ua\\
}
\maketitle
\date{\today}

\begin{abstract}
We consider a hypothetical possibility of the variability
of light velocity with time and position in space which is  derived
from two natural postulates. For the consistent consideration
of such variability we generalize translational
transformations of the Theory of Relativity.
The formulae of transformations between two rest observers
within one inertial system are obtained.
It is shown that equality of velocities of two particles is as
relative a statement as simultaneity of two events is.
We obtain the expression for the redshift of radiation of
a rest source which formally reproduces the Hubble Law.
Possible experimental implications of the theory are discussed.
\end{abstract}
\vskip2pc]

\section{\bf
                         INTRODUCTION
}

Recently a number of papers have been published
\cite{Moffat93} - \cite{Albrecht99}
in which the possibility
of light speed variability with time has been investigated.
It has been shown that models of Varying Light Speed might resolve
some cosmological problems, such as:
the flatness problem, the quasi-flatness problem \cite{Barrow99},
the horizon problem etc.

Investigations of the possibility of variability of fundamental
constants with time have a long history and
there are various approaches to the problem \cite{Dirac37}-\cite{Gammow48}. 
After it had become clear that such a fundamental constant as the
curvature radius
of our Universe varies with time, a doubt arose about
the constancy of other physical constants.
An excellent review of the research devoted to the variability of physical
constants with time can be found in Ref.~\cite{Chechev78}.

It is obvious that introduction of variability
of light speed with time is not
possible without considerable modification of the Theory of Relativity.
Recently a generalization of the Lorentz transformation
(so-called Projective Lorentz Transformation), has been obtained 
\cite{Manida99}, \cite{Steps99}.
Within this approach the variability of light speed with
time and distance arises naturally from the analysis of transformations
between two observers within different inertial reference systems.
In the Projective Theory of Relativity, besides
the fundamental speed $c$ there
exists a new constant $\lambda$ that determines the magnitude
of corrections to the Theory of Relativity.
If $\lambda=0$, we return to the Lorentz transformations
and the Theory of Relativity.

In this paper we consider in detail a possibility
of the variability of light speed
with time, and show that to describe it consistently
it is necessary
to modify not only the Lorentz transformation
but also the translational transformations
between two rest observers.
This eliminates some contradictions and makes the physical picture
clearer.

In Sec. II, on the basis of two simple postulates we obtain
the form of a functional dependence of light speed on time and distance.
Agreement between
the variability of light speed and the relativistic principles
requires modification
of the Theory of Relativity, which is considered in Sec. III
for the case of rest
observers within one inertial reference system.
It is shown in Sec. IV that
the  variability of the light speed with time
and distance results in Hubble's redshift
for rest sources.  The formulae for
aberration are obtained which can also be interpreted in
terms of Hubble Law. Possible
experimental implications of the theory are studied in Sec. V.

\section{\bf
                A TIME-SPACE VARYING SPEED OF LIGHT
}

First of all  let us consider in general the possibility of the variability
of light speed with time.
Our purpose is to obtain most simple and natural mode of
variability of speed with
time. In particular, it is preferable that photons
\footnote{
By "photon" we mean a light signal or wave packet that is much
smaller than the distance it is travelling}
still move without acceleration.

We require that the following postulates hold:
\vskip 0.5 cm

{\sl
1. The Light Speed varies with time and distance: $C \to C(t,\vec{r})$
}
\vskip 0.1 cm

{\sl
2. The speed of a particular photon is constant along its trajectory.
}

\vskip 0.5 cm

The first postulate seems obvious from the relativistic point of view.
If a physical constant varies with time it must vary with distance as well.
The second one in some respect introduces the variability of light
speed with time minimally. This means that though in some point of space
$r_0$ the light speed varies with time $C(t,r_0)$,
if we observe the movement of
the particular photon, we will find it travelling uniformly
along the trajectory
$\vec{r}=\vec{r}_0+\vec{C}(t_0,\vec{r}_0)(t-t_0)$,
at a constant
speed $\vec{C}_0=\vec{C}(t_0,\vec{r}_0)=\vec{C}(t,\vec{r})$,
where $\vec{r}_0$, $t_0$ are some fixed point and moment of time.
In other words,
the function of light speed $\vec{C}(t,\vec{r})$ must satisfy
the following functional equation:
\begin{equation}\label{FunctC}
 \vec{C}\left(t,\vec{r}_0+\vec{C}(t_0,\vec{r}_0)(t-t_0)\right)
 =\vec{C}\left(t_0,\vec{r}_0\right)
\end{equation}
for any $t$, $t_0$, $\vec{r}_0$.

To solve of this equation, let us
consider the trajectory of the moving photon .
Since $\vec{C}_0=\vec{C}(t_0,\vec{r}_0)$,
      $\vec{r}_0$ is a function of $\vec{C}_0$ and $t_0$.
Thus the trajectory of the photon
\begin{equation}
  \vec{r}=\vec{r}_0+\vec{C}_0 \cdot (t-t_0)=\vec{F}_1(\vec{C}_0,t_0)+\vec{C}_0t
\end{equation}
or, since $\vec{C}_0=\vec{C}(t,\vec{r})$ we have
\begin{equation}
  \vec{F}_1\left(\vec{C}(t,\vec{r}),t_0\right)=\vec{r}-\vec{C}(t,\vec{r})t.
\end{equation}
The fixed moment of time $t_0$ can be chosen arbitrarily
and does not depend on
the current position $\vec{r}$ and time $t$, thus the function
$\vec{F}_1$ does not depend on $t_0$.
So, the most general solution of the equation (\ref{FunctC})
has the following form:
\begin{equation}\label{SolFunctC}
 \vec{C}(t,\vec{r})=\vec{F}\left(\vec{r}-\vec{C}(t,\vec{r})t\right)
\end{equation}
where $\vec{F}(\vec{\xi})$ is an arbitrary function.

To make the function $\vec{F}(\vec{\xi})$ a more specific one
we need to introduce additional postulates.
It is however easy to see that
resolving Eq.(\ref{SolFunctC}) in elementary functions is only possible
if $\vec{F}$ is linear:
$\vec{F}(\vec{\xi})=\vec{c}+\lambda c^2 \vec{\xi}$, where
$\vec{c}$ and $\lambda c^2$ are constants. Therefore, the simplest
non-trivial dependence of the light speed on time and distance satisfying the
above formulated axioms has the following form:
\begin{equation}\label{Ctr}
 \vec{C}(t,\vec{r})=
 \frac{\vec{c}+\lambda c^2 \vec{r}}{1+\lambda c^2 t}
\end{equation}

The constant $\lambda$ is a new fundamental constant which
determines the magnitude of effects
caused by dependence of light speed on time
and distance. In particular,
if $\lambda=0$, the light speed is constant and is equal to the constant
$c$ ($\vec{c}=c~\vec{n}$, where
$\vec{n}$ is a unit vector).
The initial moment of time $t=0$ corresponds to the present moment
when the fixation of units of measurement takes place. The unit
of time is chosen so that the light velocity is equal to
$C(0,0)=c=299792458~m~s^{-1}$ at that moment ($t=0$).

If the parameter $\lambda$ is small,
the effects connected with the light speed
variability with time and distance will manifest themselves in long times $t$
and at big distances $r$ from an observer. That is, only at
cosmological scale.

As it was mentioned in Sec.I, consistent introduction of light speed
variability with time and distance requires a considerable generalization of
the Theory of Relativity. In Refs. \cite{Manida99},\cite{Steps99} it
was shown how such a generalization can be applied to
the Lorentz transformation.

The functional dependence (\ref{Ctr}) requires
a generalization of transformation between two rest observers within
one inertial reference system. Indeed, let us consider the rest observer
at the origin $x=0$ which at the moment $t=0$ emits a light signal
in the direction of the second rest observer at the point $x=R$.
The speed of this signal equals $C(0,0)=c$, and propagating according to
the second postulate at the constant speed $c=C(t,ct)$, it
reaches the second observer
at the moment of time $t=R/c$. However, the second observer
cannot reflect this
signal with the same speed because in that case it would return to $x=0$
with the speed "$c$" which is greater than the speed of light for that moment:
\begin{equation}
c > C\left(\frac{2R}{c},0\right)=\frac{c}{1+2 \lambda c R}.
\end{equation}
It is especially strange from
the point of
view of the observer at $x=0$, because for him the light speed
\begin{equation}
\vec{C}(t,0)=\frac{\vec{c}}{1+\lambda c^2 t}
\end{equation}
is isotropic, and he can
receive and emit signals with the same speed in any direction ( for the given
moment of time).

Such a seeming non-equality of two rest observers shows that it is
necessary to consider in detail
the relation not only between the measurements performed by
observers in different inertial frames of reference but also between
observers within the same inertial frame.
These transformations along with the Projective Lorentz Transformations
\cite{Manida99},\cite{Steps99} provide the necessary
generalization of the Theory of Relativity.

\section{\bf
            GENERALIZATION OF TRANSLATIONAL TRANSFORMATIONS.
}

Let us consider two rest observers within one inertial system
who are situated at the points $x=0$ and $x=R$.
We denote coordinates and times of events as
measured by the first and second observers respectively by
$x,y,z,t$ and $X,Y,Z,T$. The question is as follows:
"What is the most natural way to generalize translational transformations?"

\begin{equation}
  \left\{\begin{array}{l}
     X=x-R,\\
	 Y=y \\
	 T=t
  \end{array}\right.
       \stackrel{?}{\longmapsto}~~~
  \left\{\begin{array}{l}
	 X=X(x,y)\\
	 Y=Y(x,y) \\
	 T=T(x,y,t)
  \end{array}\right.
\end{equation}
(Below we will only consider two dimensions $(x,y)$, because all the formulae
for $y$ and $z$ components are  equivalent.)

To solve the stated problem we use the Principle of
Parametrical Incompleteness
\cite{Steps99}
which consists in the following.
The set of axioms of classical mechanics is complete
and any statement formulated within the theory framework
can be either proved or
denied on the basis of these axioms.
Reducing the  number of axioms would result in appearance of
indeterminable parameters and functions, i. e. incompleteness
of the theory.
However, there possibly are such informational simplifications
that only a finite set of constants remain indeterminable.
These constants then will play the role of the fundamental physical
constants and incompleteness will be parametrical.

In this way one could build the relativistic theory
with the constant $c$ and quantum mechanics with the
Planck constant $\hbar$.
This is, so to say, the principle of correspondence inversely.
We conventionally obtain classical mechanics from relativistic mechanics
in the limit $c=\infty$.
However, it is possible to obtain
relativistic mechanics (and other generalizations)
from classical mechanics, by reducing the number of axioms.
With each of these generalizations of classical mechanics
some fundamental physical constant will be connected.

Let us formulate five axioms concerning two observers in the same
reference frame.
\begin{center}\bf
                      Axioms
\end{center}

{\sl
1. The transformations of coordinates and time are continuous,
   differentiable and single valued-functions.
}

{\sl
2. If from the point of view of
   one observer a free particle moves uniformly,  it will move
   uniformly from the point of view of another observer.
}

{\sl
3. The observers negotiate a units of length
   so that their relative distance is equal to $R$.
}

{\sl
4. All the observers are equal and the transformations compose a group.
}

{\sl
5. Space is isotropic.
}
\vskip 0.2cm

The first axiom is standard for the majority of physical constructions.
The second one is actually a definition of inertial reference systems and
time. We define the time so that the movement of a free particle
is as simple as possible.
The third one is a definition of units of length:
two rest observers, assume, by mutual agreement, that the distance
between them is equal to $R$.
These axioms are very strong and completely fix the functional form
of transformations. We can show (see Appendix), that the most general
transformations satisfying the first three axioms are:
\begin{eqnarray}
     X&=&\frac{x-R}{1-\sigma(R) x},\nonumber\\
	 Y&=&\frac{\gamma(R) y}{1-\sigma(R) x} \label{DrobLine}\\
	 T&=&\frac{a(R) x+ b(R) t +c(R)+ d(R)y}{1-\sigma(R)x}, \nonumber
\end{eqnarray}
where $\sigma(R),~\gamma(R)~a(R),~b(R),~c(R)~,d(R)$ are some unknown functions.
The linear fractional transformations (\ref{DrobLine}) are well-known
as the most general geometrical transformations imaging
a straight line into a straight line.
This is the main point of the second axiom.

The requirement of fulfillment of group properties (axiom 4) means that there
are at least three equal observers for whom:
\begin{equation}
x_2=\frac{x_1-R_1}{1-\sigma_1 x_1},~~~
x_3=\frac{x_2-R_2}{1-\sigma_2 x_2}=\frac{x_1-R_3}{1-\sigma_3 x_1},
\end{equation}
where $\sigma_i=\sigma(R_i)$. These equations are satisfied only if
\begin{equation}
\frac{\sigma(R_1)}{R_1}=\frac{\sigma(R_2)}{R_2}=\alpha=const
\end{equation}
and
\begin{equation}
R_3=\frac{R_1+R_2}{1+\alpha R_1R_2}.
\end{equation}
Since relative distances $R_1$ and  $R_2$ are arbitrary,
$\alpha$ is a fundamental constant which is the same for all the observers.

The reverse transformation corresponds to substitution $R\to -R$, and
since
\begin{equation}
y=\frac{1-\alpha R^2}{\gamma(R)}\frac{Y}{1+\alpha R X},
\end{equation}
we have $\gamma(R)\gamma(-R)=1-\alpha R^2$.
The isotropy of space (axiom 5) implies that transformations are invariant
under inversion of the spatial axes $y\to -y$, $Y\to -Y$, $R\to -R$ etc.
This leads to the fact that the function $\gamma(R)$ is even,
and for the space transformations we obtain
\begin{equation}\label{SpaceTrun0}
     X=\frac{x-R}{1-\alpha R x},~~~
     Y=\frac{y\sqrt{1-\alpha R^2}}{1-\alpha R x}.
\end{equation} .

These formulae formally coincide with
the velocity transformations
in the relativistic theory. It means that observers are placed
in homogeneous and isotropic space of constant curvature.
The coordinates they
use to measure physical distance are Cartesian coordinates on Beltrami's
map. Beltrami's space touches the space at the point where the observer is
situated, and it possesses the property that any geodesic line is projected
on it as a straight line. In the simplest
case of a two-dimensional sphere, Beltrami's map
is a plane tangent to the sphere.
The projection on the plane is made from the
centre of the sphere.
Physical and geometrical distances to some point are connected by the
equation $S_{phys}=\tan(S_{geom})$, and for Lobachevsky
space of negative curvature by
$S_{phys}=\tanh(S_{geom})$. Analogous relations between
geometrical and physical values also exist in the velocity space
of the Theory of Relativity.

Now let us consider the transformation of time.
Suppose that all events lying
in some plane normal to the $x$ axis occur simultaneously
from the point of view of one observer. Then they will be simultaneous
from the point of view of another observer as well.
It means that $T=T(x,t)$ and $d(R)=0$.
The requirement of isotropy
(axiom 5) leads to the fact that the functions $c(R),b(R)$
are even, and $a(R)$ is an odd one.
Analogously to the coordinate case we find
the reverse transformation and require it to coincide with the initial one
after the replacement $R\to -R$. This gives the following equations:
 \begin{equation}
 b(R)=\sqrt{1-\alpha R^2},~~~
 c(R)=\frac{a(R)}{\alpha R}(\sqrt{1-\alpha R^2}-1)
\end{equation}

The composition of transformations $t_2=f(t_1,x_1,R_1)$,
$t_3=f(t_2,x_2,R_2)=f(t_1,x_1,R_3)$ is possible only if
\begin{equation}
 \frac{a(R_1)}{R_1}=\frac{a(R_2)}{R_2}=\lambda=const.
\end{equation}
So, we obtain:
\begin{equation}\label{TimeTrun0}
 T=\frac{t\sqrt{1-\alpha R^2}+\lambda R x +(\sqrt{1-\alpha R^2}-1)\lambda/\alpha}
		{1-\alpha R x}.
\end{equation}

We should note that the synchronization procedure is derived automatically:
the event that happens between observers at equal distances from these
observers,
$x=-X=(1-\sqrt{1-\alpha R^2})/\alpha R$, is simultaneous for them: $T=t$.

Using transformations (\ref{SpaceTrun0}) and  (\ref{TimeTrun0})
it is easy to obtain transformations for the speed of particle as measured
by each of the observers $\vec{U}=d\vec{X}/dT$, $\vec{u}=d\vec{x}/dt$:
\begin{eqnarray}
   U_X &=& \frac{u_x\sqrt{1-\alpha R^2}}
                {1+\lambda R u_x - \alpha R (x-u_x t)} \label{SpeedX}\\
   U_Y &=& \frac{u_y+\alpha R (yu_x-xu_y)}
                {1+\lambda R u_x - \alpha R (x-u_x t)}. \label{SpeedY}
\end{eqnarray}
If the particle moves uniformly $\vec{r}=\vec{r}_0+\vec{u}t$,
transformations of speed do not vary with time but vary with
the "initial" position of the particle $\vec{r}_0$.

Here we should note that, if $\alpha=(\lambda c)^2$,
the formula (\ref{Ctr}) for the light speed $\vec{C}(t,\vec{r})$
possesses the following properties:

1. $\vec{C}(t,\vec{r})$ is invariant for both observers.
It means that, if $\vec{C}(t,\vec{r})$ is transformed as a speed
(\ref{SpeedX}),(\ref{SpeedY}), the same function expressed in coordinates
of each observer stands on the right and on the left
of the transformations (\ref{SpeedX}),(\ref{SpeedY}).
In case of light moving along the $x$ axis we have:
\begin{equation}
   C(T,X) = \frac{ C(t,x)\sqrt{1-(\lambda c R)^2}}
            {1+\lambda R  C(t,x) - (\lambda c)^2 R (x-C(t,x)t)}
\end{equation}
The movement in an arbitrary direction is considered in the next
section.

2.  $\vec{C}(t,\vec{r})$ is the maximal possible speed 
for the given point of space $\vec{r}$ and given moment of time $t$.

On the basis of these two properties we call $\vec{C}(t,\vec{r})$
the speed of light.

Therefore, for the consistent introduction 
of the varying with time and distance
light velocity (\ref{Ctr}) into the theory, it is necessary
to generalize the translational transformations for rest observers:
\begin{equation}\label{SpaceTrun}
     X=\frac{x-R}{1-(\lambda c)^2 R x},~~~
     Y=\frac{y\sqrt{1-(\lambda c R)^2}}{1-(\lambda c)^2 R x}.
\end{equation} .

\begin{equation}\label{TimeTrun}
 1+\lambda c^2 T=\frac{\sqrt{1-(\lambda c R)^2}}{1-(\lambda c)^2 R x}
                           \left(1+\lambda c^2 t\right).
\end{equation}
If two observers move at a relative speed $v$, the generalized
Lorentz transformations have the following form
\cite{Manida99},\cite{Steps99}:
\begin{eqnarray}
    x'&=&\frac{\gamma(x-vt)}{1+\lambda v \gamma x-\lambda c^2 (\gamma-1)t}, \label{PLTx}\\
    y'&=&\frac{y}{1+\lambda v \gamma x-\lambda c^2 (\gamma-1)t},            \label{PLTy}\\
    t'&=&\frac{\gamma(t-vx/c^2)}{1+\lambda v \gamma x-\lambda c^2 (\gamma-1)t},	\label{PLTt}
\end{eqnarray}
where $\gamma=1/\sqrt{1-v^2/c^2}$  is the Lorentz factor.
The formulae (\ref{SpaceTrun})-(\ref{PLTt}) form the basis
of kinematics of the Projective Theory of
Relativity, within which the speed of light varies with time but at the
same time is an invariant of the theory.

\vskip 0.5cm

The contradiction considered in the Sec. II is easy to resolve now. From the
point of view of the first observer the signal reaches the second
observer $x=R$ at the moment of time $t=R/c$ with the speed $u=c$.
From the point of view of the second observer the speed of the signal
(\ref{SpeedX})
and the moment of time (\ref{TimeTrun}) are equal to:
\begin{equation}
 U=c\sqrt{\frac{1-\lambda c R}{1+\lambda c R}},~~
 \lambda c^2 T = \sqrt{\frac{1+\lambda c R}{1-\lambda c R}}-1.
\end{equation}
The observer reflects the signal with the same speed $U\to -U$.
However, due to the transformations of speed its speed relative
to the first observer (\ref{SpeedX}) equals:
\begin{equation}
 u=-c\frac{1-\lambda c R}{1+\lambda c R}.
\end{equation}
In a time $t=R/c+R/|u|$ the signal returns to the first observer to the point
$x=0$, and has the same speed as any other light signal at that moment of time:
\begin{equation}
 C\left(\frac{R}{c}+\frac{R}{|u|},0\right)
 = c\frac{1-\lambda c R}{1+\lambda c R}=|u|
\end{equation}

Therefore, if two particles have the same speed from the point of view of
one observer, they will have different speeds for another observer.
The equality of speeds is as relative a notion
as the simultaneity of events is.
This happens because the Projective Transformations do not conserve 
parallelism of straight lines.

If we add to the initial system of axioms the requirements
of absolutivity of equality
of two speeds and absolutivity of time, we obtain the complete axiom system
in which incompleteness connected with undefineable constants
$\lambda=0$, $\alpha=0$ disappears. If we exclude these axioms, 
we obtain the more general parametrically incomplete theory with
new fundamental physical constants.
This is the Principle of Parametrical Incompleteness.

\section{\bf
  THE HUBBLE LAW. EXPANSION OF THE STATIC UNIVERSE.
}

An interesting consequence of the results of previous sections arises
when the Doppler effect is analyzed within one inertial system.

\vskip 0.3cm

{\bf 1. Hubble Low}

Let us consider a remote rest source with coordinates
$\vec{R}$ emitting light in the direction of observer which is situated at the
origin $x=0$.
According to observer's clock the light pulse emitted at the moment of time
$t_1$  reaches it at the moment $t_2$.
Since the speed of this signal is constant $C(R,t_1)=C(0,t_2)$ 
and it moves in the direction
towards the observer $\vec{c}=-c \vec{R}/R$, we have the following
relation between $R,t_1,t_2$:
\begin{equation}\label{Habbl1}
 (t_2-t_1)c = R + \lambda c^2 R t_2.
\end{equation}
Let us assume that light pulses are emitted with the period
$\tau_0=\Delta T_1$, and are received with the period $\tau=\Delta t_2$. 
Since the source's time  $T$ and the observer's
time $t$ are related by Eq. (\ref{TimeTrun}), the interval 
$\Delta T$ equals $\Delta t/\sqrt{1-(\lambda c R)^2}$ 
for $\vec{x}=\vec{R}$.
Thus the period of emission is $\tau_0=\Delta t_1/\sqrt{1-(\lambda c R)^2}$,
and introducing the parameter of redshift $z$ we finally obtain:
\begin{equation}\label{Habbl2}
  1+z=\frac{\tau}{\tau_0}=\sqrt{\frac{1+\lambda c R}{1-\lambda c R}}.
\end{equation}
Interpreting the redshift according to Doppler's formula, we obtain
the Hubble law: $ \vec{V}=\lambda c^2 \vec{R}$,
but such an interpretation would not be correct in this case.

\vskip 0.3cm
{\bf 2. Distance Measurement}

We can obtain the same result by the following speculations.
Suppose, the observer at $x=0$ makes a radiolocating experiment measuring
the distance to the rest object at $x=R$.
At the moment of time $t_1$ this observer emits a light
signal at the speed of $C(0,t_1)$ receiving it at time $t_2$
at the speed of $C(0,t_2)$.
If the observer (despite different speeds of the emitted and reflected signals)
assumed the distance to the object to be equal to $l=(t_2-t_1) c/2$, he would,
probably, conclude that the object moves away from him at Hubble's speed:
\begin{equation}\label{Habbl4}
  l=\frac{c}{2}\left(\frac{R}{C(t_1)}+\frac{R}{C(t_2)}\right)=R+\lambda c^2 R~ t
   =R+V t,
\end{equation}
where $t=(t_2+t_1)/2$. Such an interpretation would not,
of course, be correct.
If the observer emitted signals at speed $u<C(0,t_2)$, he could (with
apropriate conditions of reflection) receive them at the same speed, and the
distance $l=(t_2-t_1) u/2$ would be unchanging and equal to $R$.

\vskip 0.3cm
{\bf 3. Aberration of Light}

Let us obtain another useful equation which also can be interpreted in terms
of the Hubble speed.
An expression similar to that for aberration in the Theory of Relativity
arises for the rest light source and receiver.
Suppose, the light travels in some direction $(\cos\omega,\sin\omega)$
relative to the observer at $x=0$, and in direction $(\cos\Omega,\sin\Omega)$
relative to the observer at $x=R$. Then the
components of light velocity will be equal to
\begin{equation}\label{SpeedLightComp}
   C_x=c\frac{\cos\omega+\lambda c x}{1+\lambda c^2 t}
   \quad
   C_y=c\frac{\sin\omega+\lambda c y}{1+\lambda c^2 t}.
\end{equation}
If we put (\ref{SpeedLightComp}) and similar equations for the second observer
in (\ref{SpeedX}),(\ref{SpeedY}), where $\alpha=(\lambda c)^2$,
we would obtain the identity for any $x,y,t$, only if:
\begin{eqnarray}\label{SpeedLightOmega}
   \sin\Omega&=&\frac{\sqrt{1-(\lambda c R)^2}\sin\omega}{1+\lambda c R \cos\omega},\\
   \cos\Omega&=&\frac{cos\omega+\lambda c R}{1+\lambda c R \cos\omega}.
\end{eqnarray}
These formulae formally coincide with those
for aberration in the Theory of Relativity if we set
$\lambda c R = V/c$. So, we again come to the Hubble formula.

\vskip 0.3cm
{\bf 4. "Expansion" of the static Universe}

If we admit the possibility light speed of variability  with time,
we will necessarily come to the following cosmological model.
The Universe is a stationary space of constant curvature (the Lobachevsky
space). The curvature is not connected with the presence of matter
and is an intrinsic property of the empty space. The course of  time in
the Universe is defined so that it would look as simple as possible.
This leads to the flat pseudo-Euclidian space-time.

The evolution of the Universe is connected with the decreasing
of the speed of light
with time and 14(?) billion years ago
the speed of light was equal to infinity.
We now take this moment as the origin of time, i.e. make
the shift $t\to t-1/\lambda c^2$ \cite{Manida99}
in all the formulae .
Because of the infinite speed of interactions,
the early Universe was homogeneous and hot.
However, there was no singularity of matter.
All the clocks in the Universe were synchronized ($C=\infty$) and
pointed at the zero time mark:
\begin{equation}
T=\frac{\sqrt{1-(\lambda c R)^2}}{1-(\lambda c)^2 \vec{R} \vec{r}}t.
\end{equation}

With the course of time the speed of light was decreasing,
the Universe was cooling, and the clocks located at the
distance $r=R$ from us started to advance compared to our clock:
\begin{equation}
T=\frac{t}{\sqrt{1-(\lambda c R)^2}}>t
\end{equation}
Nevertheless, we observe the Universe in its past state
\begin{equation}
T_v=\sqrt{\frac{1-\lambda c R}{1+\lambda c R}}t=\frac{t}{1+z} < t,
\end{equation}
because the speed of light is finite $C(t,0)=(\lambda c t)^{-1}$
(here $z$ is the parameter of redshift).

The frequency of the light we receive from remote rest sources is
shifted to the red. The farther the source is situated
from us the more
the frequency of the light is shifted to the red,
in agreement with the Hubble Law.

The distance $R_m = 1/\lambda c$ is the maximal possible distance 
an observer can measure, 
and at the same time is the radius of curvature
of the Lobachevsky space.
\footnote{
We  point out that we are talking here about physical distances but not
about geometrical distances which are unlimited in the Lobachevsky space.
The situation is completely identical to the velocity space of
the theory of relativity, for which there is the maximum possible speed
$c$ but there is no finite limit on geometrical distance
$s= {\rm artanh}~ (u/c)$.}
At any moment of time according to our clocks $t$ we see areas situated
at the distance $R_m$ from us at the moment of time $T=0$ according
to the local clock.
The infinite value of the red shift parameter $z$ corresponds to
these areas.

Although the Hubble Law is realized automatically in this cosmological
model, it is
obvious that including matter and gravitation into consideration
can change the properties of our Space in some way, for instance,
to make it expand. In this case the Hubble effect will
consist of two components - the usual Doppler redshift
and the shift connected
with the new fundamental constant $\lambda$. As a result,
the actual age of our Universe  could be much greater than the value
derived from the Hubble Law.


\section{\bf
          CONCLUSION: VARYING SPEED OF LIGHT AND EXPERIMENT
}

Let us discuss applicability of the proposed theory
to the real World.
Since Hubble's effect is naturally described within
the Projective Theory of Relativity,
it would be interesting to associate Hubble's constant
$H=65~km/sec/Mps=6.7~10^{-11}~year^{-1}$
with the constant $\lambda c^2$.
In this case the change of the light velocity with time
would be as follows ($r=0$, $t=0$ ):
\begin{equation}\label{SpeedLightChange}
  \frac{\Delta C}{C}=-\lambda c^2 \Delta t = - 6.7~10^{-11} \frac{\Delta t}{year}.
\end{equation}

Obviously, the dimensional value $C(t,0)$
can be expressed in terms of some units of length and time,
e.g. the atomic units
$\hbar^2/m e^2$ and $\hbar^3/m e^4$.
In particular, the dimensionless combination
$\alpha(t)=e^2/\hbar C(t)$ should change.
The laboratory value of $\alpha$ is known at present (1997)
with accuracy of $4~10^{-9}$:
$\alpha^{-1}=137.03599993(52)$, which is close enough to change
(\ref{SpeedLightChange}).

Here let us make clear one point about testing the dependence of
the light velocity on time. There are two entities in our theory:
$C(t,\vec{r})$ and $c$.
The first one is the  light velocity and the maximal possible
speed of material objects,
the second one is the fundamental speed arising from
the parametrical incompleteness of the axioms of the theory.
Only after generalization
of Quantum Electrodynamics for the case of
the Projective Theory of Relativity
would it be possible to say
which of the constants would enter into $\alpha$.
The fine-structure constant   may, thus,  depend on $c$:
$\alpha=e^2/\hbar c$, and do not change with time.

Recently, a new direct limit on
$|\dot{\alpha}/\alpha|<10^{-14} year^{-1}$,
has been derived from spectral properties
of distant ($z=1\div 3.5$) quasars \cite{ExpFineStr1},\cite{ExpFineStr2}.
However, this does not mean that  (\ref{SpeedLightChange}) is falsified by
experiment.
Indeed, we observe an object which is situated at the distance $R$ from us
in its past state  at the moment $t=-R/c$
according to our clock. That time
corresponds to the local time of an object
$T=-z/(1+z)\lambda c^2$
and, therefore, the light velocity measured
by the observer, which is situated near the object, equals
$C(0,T)=c(1+z)$.
From his point of view, the dimensionless combination
$\alpha(T)=e^2/\hbar C(0,T)$,
is $1+z$ times less than our measurement shows at
the present moment of time $t=0$.
According to the rule of transformation for speeds measured by distant
observers, the light emitted by the object at the speed of $c(1+z)$
is equal to $c$
from our point of view  and,
therefore, $\alpha=e^2/\hbar c$. That is why the measurement
of $\alpha$ based on the spectra of quasars does
not allow us to test the change of light velocity in time.

It is likely that only direct laboratory
measurement of the light velocity in terms of the atomic units of length
and  time would provide a direct test for (\ref{SpeedLightChange}).

\vskip 1 cm
{\bf ACKNOWLEDGMENTS }
\vskip 0.2 cm

I would like to thank Prof. Orlyanskij and Prof. Manida for
fruitful discussions and Dr. Zaslavsky and Dr. Tishchenko
for their comments on this manuscript.

\vskip 1 cm
{\bf APPENDIX }
\vskip 0.2 cm

Let us consider arbitrary independent differentiable transformations of
the coordinate $x$ and time $t$:
\begin{equation}\label{Primxtx't'}
  X=f(x),  \quad  T=g(x,t).
\end{equation}
We require the system of coordinates $(x,t)$ and $(X,T)$ to satisfy the
definition of inertial reference systems:
\begin{equation}\label{PrimIner}
      \frac{du}{dt}=0 \quad \Longrightarrow \quad \frac{dU}{dT}=0,
\end{equation}
i.e. a free particle moves uniformly from the point of view of all observers.

By definition, the speeds are $u=dx/dt$
and $U=dX/dT$, thus:
\begin{equation}\label{PrimSpeed}
                 U=\frac{u f_x }{g_x u + g_t},
\end{equation}
where $g_x=\partial g(x,t)/\partial x$, etc.
Differentiating (\ref{PrimSpeed}) on T ( $dT=(g_x u + g_t)dt$ )
and taking into account that the coefficients of the obtained
polynomial in $u$ must be equal to zero (since $u$ is arbitrary)
we obtain the system of differential equations:
 \begin{eqnarray}
  f_{xx} g_x &=& g_{xx} f_x       \label{PrimSys1}\\
  f_{xx} g_t &=& 2 g_{xt} f_x     \label{PrimSys2} \\
  g_{tt} f_x &=& 0                \label{PrimSys3}.
\end{eqnarray}

Solving this system, we obtain:
\begin{eqnarray}
     f(x)&=&\frac{ax+b}{1+\alpha x},\nonumber \\
     g(x,t)&=&\frac{\gamma t+a'x+b'}{1+\alpha x}.\nonumber
\end{eqnarray}

In a more general case of two dimensions, linear fractional transformations
have the following form \cite{Fock64}:
\begin{eqnarray}
     X&=&\frac{ax+by+c}{1+\alpha x +\beta y},\nonumber \\
     Y&=&\frac{\bar{a}x+\bar{b}y+\bar{c}}{1+\alpha x +\beta y}, \label{PrimDrobLin}\\
     T&=&\frac{\gamma t+a'x+b'y+c'}{1+\alpha x +\beta y}.\nonumber
\end{eqnarray}
It is assumed that the third axiom is equivalent to the following equations:
\begin{equation}\label{PrimAxiom3}
  \left\{\begin{array}{l}
     X(0,y)=-R,~~~X(R,y)=0,\\
     Y(R,0)=0,~~~~~Y(0,0)=0.\\
  \end{array}\right.
\end{equation}
and we obtain the Eq.(\ref{DrobLine}).



\begin{thebibliography}{50}


\bibitem{Moffat93}
   J.W. Moffat,
   Int.J.Mod.Phys., {\bf 2D}, No.3, 351 (1993);
   gr-qc/9211020 (1992)

\bibitem{Clayton99}
  M.A. Clayton, J.W. Moffat
  Phys.Lett. {\bf B460}, 263-270,1999; (1999)
  astro-ph/9812481  (1998)

\bibitem{Albrecht98}
   A. Albrecht, J. Magueijo,
   Phys.Rev.,  {\bf D 59}, 43516 (1999);
   astro-ph/9811018 (1998)

\bibitem{Barrow99}
   J.D. Barrow,
   Phys.Rev., {\bf D 59}, 043515 (1999);

\bibitem{Barrow99:2}
  J.D. Barrow, J. Magueijo ,
  Class.Quant.Grav. {\bf 16}, 1435 (1999);
  astro-ph/9901049

\bibitem{Barrow99:3}
  J.D. Barrow (Sussex U.), J. Magueijo
  Phys.Lett. {\bf B447}, 246 (1999);
  astro-ph/9811073

\bibitem{Albrecht99}
    A. Albrecht,
   astro-ph/9904185 (1999)

\bibitem{Dirac37}
  P.A.M.Dirac,
  Nature {\bf 139}, 323 (1937);
  Proc.Roy.Soc.London, {\bf A165}, 198 (1938)

\bibitem{Teller48}
   E. Teller, Phys.Rev. {\bf 73}, 801 (1948)

\bibitem{Jordan55}
   P. Jordan,
   {\sl Schw\-erckraft und weltall}, Braunschweig (1955)

\bibitem{Brans61}
   C. Brans, R.H. Dicke,
   Phys.Rev. {\bf 124}, 925 (1961)


\bibitem{Gammow48}
   G. Gamow,
   Phys.Rev.Lett. {\bf 19}, 759 (1967)

\bibitem{Chechev78}
   V.P. Chechev, Ya.M. Kramarovsky,
   {\sl Radioactivity and Evolution of Universe} (in russian), Nauka (1978)


\bibitem{Manida99}
   S.N. Manida,
   gr-qc/9905046 (1999)

\bibitem{Steps99}
   S.S. Stepanov,
   physics/9909009 (1999)

\bibitem{ExpFineStr1}
   J.K. Webb, V.V. Flambaum, C.W. Churchill, M.J. Drinkwater, J.D Barrow,
   astro-ph/9803165 (1998)

\bibitem{ExpFineStr2}
   A.V. Ivanchik,  A.Y. Potekhin,  D.A. Varshalovich,
   astro-ph/9810166 (1998)

\bibitem{Fock64}
   V.A. Fock, {\sl The theory of space, time and gravitation},
   Pergamon Press (1964)

\end{thebibliography}
\end{document}